\documentstyle [12pt] {article}
\title{$C^{\frac {1}{2}}P^{\frac {1}{2}}T^{\frac {1}{2}}$ SYMMETRY OF THE CHAOTIC INFLATION DYNAMICS }
\author{Vladan Pankovi\'c$^{\ast,\sharp}$, Jovan Ivanovi\'c$^\sharp$ \\
$^\ast$Department of Physics, Faculty of Sciences\\
21000 Novi Sad, Trg Dositeja Obradovi\'ca 4, Serbia\\
$^\sharp$Gimnazija, 22320 Indjija, Trg Slobode 2a, Serbia\\
vdpan@neobee.net}

\date {}
\begin{document}
\maketitle

\vspace {0.5cm}


\vspace{1cm}

\begin {abstract}
In this work is shown that standard dynamical equations of the
chaotic inflation for a massive real scalar field hold an especial
local gauge symmetry. Given symmetry admits that inflation field
mass can be considered as an especial "charge". Also, it is shown
that given equations except expected $CPT$ symmetry, hold
surprisingly, $C^{\frac {1}{2}}P^{\frac {1}{2}}T^{\frac {1}{2}}$
symmetry too.
\end {abstract}
\vspace {1.5cm} Key words: chaotic inflation - massive scalar
field - $CPT$ symmetry \vspace {1.5cm}

In this work will be shown that standard dynamical equations of
the chaotic inflation for a massive real scalar field hold an
especial, local gauge symmetry. Given symmetry admits that
inflation field mass can be considered as an especial "charge".
Also, it is shown that given equations except $CPT$ symmetry, hold
$C^{\frac {1}{2}}P^{\frac {1}{2}}T^{\frac {1}{2}}$ symmetry too.

As it is well-known [1]-[7], chaotic inflation dynamics for a real
scalar field with mass $m$ can be generally presented by two usual
differential equations
\begin {equation}
   \frac {d^{2}\phi}{dt^{2}}+ 3 \frac{1}{a} \frac {da}{dt} \frac{d\phi}{dt} = - m^{2}\phi
\end {equation}
\begin {equation}
   (\frac{1}{a} \frac {da}{dt})^{2} + \frac {k}{{\it a}^{2}} = \frac {4\pi}{3m^{2}_{P}}((\frac {d\phi}{dt})^{2} + m^{2}\phi^{2}) .
\end {equation}
Here $\phi$ represents the real scalar field depending of the time
$t$, $m_{P}$ - Planck's mass, ${\it a}$ - scale factor of the
universe, $k$ - curvature constant (that equals 1 for closed, 0
for flat and -1 for open universe).

Define the following operator
\begin {equation}
   D_{t} = \frac {d}{dt} - im
\end {equation}
representing an especial covariant differentiation so that it is
satisfied
\begin {equation}
   D_{t}(e^{imt} \phi) = e^{imt} \frac {d\phi}{dt}   .
\end {equation}

By use of (3), (4) equation (1) can be presented in an equivalent
form
\begin {equation}
  D_{t} D_{t}(e^{imt} \phi) + 3\frac{1}{a} \frac {da}{dt}D_{t} (e^{imt} \phi) = - m^{2}(e^{imt} \phi)
\end {equation}
or after complex conjugation, $*$, in other equivalent form
\begin {equation}
  D^{*}_{t} D^{*}_{t}(e^{-imt} \phi) + 3\frac{1}{a}\frac {da}{dt} D^{*}_{t} (e^{-imt} \phi) = - m^{2}(e^{-imt} \phi)
\end {equation}
Also, by use (3), (4) equation (2) can be presented in an
equivalent form
\begin {equation}
(\frac{1}{a} \frac {da}{dt})^{2} + \frac {k}{a^{2}} =
\frac{4\pi}{3m^{2}_{P}}( D_{t} (e^{imt} \phi) D^{*}_{t}
(e^{-imt}\phi) + m^{2}(e^{imt} \phi) (e^{-imt} \phi))
\end {equation}
that does not change by complex conjugation.

It is not hard to see that equations (1), (2) and (5), (7) (or
(6), (7) ) are not only equivalent but they have equivalent forms.
It simply means that dynamics of the chaotic inflation with
massive scalar field is invariant, i.e. symmetric in respect to an
especial local gauge transformation of the scalar field
\begin {equation}
  \phi \rightarrow e^{imt} \phi
\end {equation}
or complex conjugated transformation
\begin {equation}
  \phi \rightarrow e^{-imt} \phi
\end {equation}

It can be observed that transformation
\begin {equation}
   m \rightarrow -m
\end {equation}
does not change (1) and (2) as well as (7). On the other hand the
same transformation changes (5) in (6) and vice versa. Generally
speaking dynamics of the chaotic inflation with massive scalar
field is invariant, i.e. symmetric in respect to transformation
(10). But, according to (8), (9) it is obvious that transformation
(10) can be compensated by complex conjugation, or roughly
speaking, that (10) corresponds to complex conjugation. It opens
an interesting possibility. Namely, according to characteristics
of charge conjugation transformation, $C$, in quantum field theory
[8], [9], mass of the inflationary quantum field can be
effectively treated as a "charge". Simultaneously, symmetry of the
dynamics of the chaotic inflation with massive scalar field in
respect to (10) can be effectively considered as an especial type
$C$ symmetry. In this sense dynamics of the chaotic inflation with
massive scalar field is $C$ symmetric.

Further, time reversal transformation, $T$, [8], [9],
\begin {equation}
   t \rightarrow -t
\end {equation}
does not change (1) and (2) as well as (7). On the other hand the
same transformation changes (5) in (6) and vice versa. It means
that dynamics of the chaotic inflation with massive scalar field
is $T$ symmetric.

Finally, as it is well-known, demand for homogeneity of the space
determine Robertson-Walker line element, which in the spherical
coordinates, has form
\begin {equation}
  ds^{2} = -dt^{2} + {\it a}^{2} (\frac {dr^{2}}{1-kr^{2}} + r^{2}d\Omega^{2})        .
\end {equation}
where
\begin {equation}
  r^{2} = x^{2}+ y^{2}+ z^{2}.
\end {equation}
Obviously, parity transformation, $P$, [8], [9],
\begin {equation}
   x \rightarrow -x
\end {equation}
\begin {equation}
   y \rightarrow -y
\end {equation}
\begin {equation}
   z \rightarrow -z
\end {equation}
does not change (13) and (12) as well as dynamics of the chaotic
inflation with massive scalar field (1), (2) or (5), (7) (or (6),
(7) ).

In this way it is shown that dynamics of the chaotic inflation
with massive scalar field is $C$ symmetric, $P$ symmetric and $T$
symmetric as well as $CP$, $CT$, $PT$ and $CPT$ symmetric.

Define additionally the following three transformations
\begin {equation}
   m \rightarrow im
\end {equation}
\begin {equation}
   t \rightarrow it
\end {equation}
\begin {equation}
   k \rightarrow -k
\end {equation}
where $i$ represents the imaginary unit $(-1)^{\frac {1}{2}}$.

According to (10) and previous discussion (17) can be considered
as $C^{\frac {1}{2}}$ transformation.

According to (11) transformation (18) can be considered as
$T^{\frac {1}{2}}$ transformation.

 Application of (19) and the following transformation
\begin {equation}
   r \rightarrow ir
\end {equation}
i.e.
\begin {equation}
   x \rightarrow ix
\end {equation}
\begin {equation}
   y \rightarrow iy
\end {equation}
\begin {equation}
  z \rightarrow iz
\end {equation}
representing, obviously, $P^{\frac {1}{2}}$, at ${\it a}^{2}(\frac
{dr^{2}}{1-kr^{2}} + r^{2}d\Omega^{2})$ changes given expression
in an equivalent expression $-{\it a}^{2}(\frac {dr^{2}}{1-kr^{2}}
+ r^{2}d\Omega^{2})$. In this way transformation (19) can be
compensated by transformations (21)-(23), i.e. $P^{\frac {1}{2}}$,
or, roughly speaking, transformation (19) can be considered
equivalent to transformations $P^{\frac {1}{2}}$.

It is not hard to see that none of the equations (1)-(7) is
invariant, i.e. symmetric in respect to $C^{\frac {1}{2}}$,
$P^{\frac {1}{2}}$, $T^{\frac {1}{2}}$, $C^{\frac {1}{2}}P^{\frac
{1}{2}}$, $C^{\frac {1}{2}}T^{\frac {1}{2}}$ and $P^{\frac
{1}{2}}T^{\frac {1}{2}}$. Nevertheless, as it is not hard to see
too, any of the equations (1)-(7) is invariant, i.e. symmetric in
respect to $C^{\frac {1}{2}}P^{\frac {1}{2}}T^{\frac {1}{2}}$. In
this way it is shown that dynamics of the chaotic inflation with
massive scalar field is $C^{\frac {1}{2}}P^{\frac {1}{2}}T^{\frac
{1}{2}}$ symmetric.

In conclusion it can be shortly repeated and pointed out the
following. In this work it is shown that standard dynamical
equations of the chaotic inflation for a massive real scalar field
hold an especial, local gauge symmetry. Given symmetry admits that
inflation field mass be considered as an especial "charge". Also,
it is shown that given equations except expected $CPT$ symmetry,
hold, surprisingly, $C^{\frac {1}{2}}P^{\frac {1}{2}}T^{\frac
{1}{2}}$ symmetry too. All this represent an interesting result
whose detailed discussion goes over basic intentions of this work.

{\bf References}

\begin {itemize}

\item [[1]] A. D. Linde, Phys.Lett., {\bf 129B}, (1983.), 177.
\item [[2]] A. D. Linde,  Mod.Phys.Lett., {\bf A1}, (1986.), 81.
\item [[3]] A. S. Goncharov, A. D. Linde, V. F. Mukhanov, J. Mod. Phys., {\bf A2}, (1987.) , 561.
\item [[4]] A. D. Linde, in,{\it  300 Years of Gravitation}, eds. S. W. Hawking, W. Israel (Cambridge University Press, Cambridge, England, 1989.)
\item [[5]] A. D. Linde, {\it Inflation and Quantum Cosmology} (Academic Press, Boston, 1990.)
\item [[6]] A. D. Linde, {\it Particle Physics and Inflationary Cosmology} (Harwood Academic Publishers, Chur, Switzerland, 1990.)
\item [[7]] A. Linde, {\it Inflationary Cosmology}, hep-th/0705.0164
\item [[8]] L. H. Ryder, {\it Quantum Field Theory}(Cambridge University Press, Cambridge, 1987.)
\item [[9]] R. F. Streater, A. S. Wightman, {\it PCT, Spin and Statistics and All That} (Addison-Wesley, New York, 1989.)

\end {itemize}

\end{document}